\documentclass[onecolumn,amsmath,amssymb,floatfix,nofootinbib]{revtex4}

\usepackage{graphicx}
\usepackage{dcolumn}
\usepackage{bm}

\def\VEV#1{\left\langle #1 \right\rangle}
\newcommand{\delt}{{\vec\nabla_{\vec\theta}}}
\def\be{\begin{equation}}
\def\ee{\end{equation}}

\def\cleb#1#2#3#4#5#6{\langle #3 \, #4 \, #5 \, #6 | #1#2 \rangle}
\def\ALMt#1#2#3#4{A^{#1 #2}_{#3 #4}}
\def\ALM{\ALMt{L}{M}{l}{l'}}
\def\ALMp{{A^{\oplus}}^{LM}_{ll'}}
\def\ALmMp{{A^{\oplus}}^{L \,-M}_{ll'}}
\def\ALMm{{A^{\ominus}}^{LM}_{ll'}}
\def\ALmMm{{A^{\ominus}}^{L \, -M}_{ll'}}

\begin{document}

\title{Odd-Parity Bipolar Spherical Harmonics}

\author{Laura G. Book$^1$, Marc Kamionkowski$^{12}$, and Tarun Souradeep$^3$}
\affiliation{$^1$California Institute of Technology, Mail Code 350-17,
     Pasadena, CA 91125}
\affiliation{$^2$Johns Hopkins University, Department of Physics and Astronomy, Bloomberg 439, 3400 N. Charles St., Baltimore, MD 21218}
\affiliation{$^3$Inter-University Centre for Astronomy and
     Astrophysics, Pune 411007, India} 

\date{\today}

\begin{abstract}
Bipolar spherical harmonics (BiPoSHs) provide a general formalism for
quantifying departures in the cosmic microwave background (CMB) from
statistical isotropy (SI) and from Gaussianity.  However, prior work
has focused only on BiPoSHs with even parity.  Here we show that there
is another set of BiPoSHs with odd parity, and we explore their
cosmological applications.  We describe systematic artifacts in
a CMB map that could be sought by measurement of these
odd-parity BiPoSH modes. These BiPoSH modes may also be produced
cosmologically through lensing by gravitational waves (GWs),
among other sources. We derive expressions for the BiPoSH modes
induced by the weak lensing of both scalar and tensor
perturbations. We then investigate the
possibility of detecting parity-breaking physics, such as chiral
GWs, by cross-correlating opposite parity BiPoSH modes with multipole
moments of the CMB polarization.  We find that the expected
signal-to-noise of such a detection is modest.
\end{abstract}


\maketitle

\section{Introduction}

The detection of anisotropies in the cosmic microwave background (CMB)
\cite{Smoot:1992td} has revolutionized the precision with which
cosmological measurements can be made. Most of the information that
has been obtained from the CMB so far has come from its power
spectrum, the two-point correlation function, under the assumptions of
isotropy and homogeneity. However, in recent years, attention has been
paid to effects that go beyond the power spectrum, such as weak
lensing \cite{Lewis:2006fu}, cosmic birefringence
\cite{Carroll:1989vb,Caldwell:2011pu}, and departures from statistical
isotropy (SI) \cite{de OliveiraCosta:2003pu, Hansen:2004vq,
Hanson:2009gu} and from Gaussianity \cite{Verde:1999ij,
Komatsu:2001rj, Kunz:2001ym}.

Bipolar spherical harmonics (BiPoSHs)
\cite{Hajian:2003,Hajian:2005jh,Joshi:2009mj} provide an elegant and general
formalism for quantifying a number of these physical effects.  If the
CMB map is Gaussian and statistically isotropic, then its statistics
are specified entirely in terms of the power spectrum $C_l$, the
expectation value of the squared magnitude of the spherical-harmonic
coefficients $a_{lm}$ for the map, and there are no
correlations between different $a_{lm}$s.  A wide variety of
departures from SI and Gaussianity induce correlations between
different $a_{lm}$s.  The point of the BiPoSH formalism is to
parametrize correlations between two different coefficients, $a_{lm}$
and $a_{l'm'}$, that represent two different ``angular-momentum''
states, in terms of total angular momenta $L$ and $M$.  Bipolar
spherical harmonics have been used to search for non-standard cosmic
topology \cite{Hajian:2003ic}, anisotropy in primordial power
\cite{Pullen:2007tu,Hanson:2009gu}, and model-independent departures
from SI
\cite{Hajian:2004zn,Souradeep:2003qr,Joshi:2009mj,Souradeep:2006dz,Hajian:2006ud,Ghosh:2006xaa,Basak:2006ew}.
They have also been used to test for asymmetric beams
\cite{Hanson:2010gu} and/or other systematic artifacts in WMAP
\cite{Bennett:2010jb}.  BiPoSHs for polarization have been proposed to
search for position-dependent rotation of the CMB polarization
\cite{Kamionkowski:2008fp,Gluscevic:2009mm,Yadav:2009eb}.

However, there is still more that can be done with bipolar spherical
harmonics, and the purpose of this paper is to enumerate some of them.
First and foremost, we point out here that almost all prior work on
BiPoSHs has considered only BiPoSHs with even parity
(Sec.~\ref{sec:review}).  There exists an entire other set of BiPoSHs
that have the opposite parity, and these can provide probes of both
cosmological effects and systematic artifacts that would remain
elusive with the even-parity BiPoSHs that have been considered so far.
We show, for example, that lensing by gravitational waves (GWs) can
excite odd-parity BiPoSHs, and we describe a pointing error that could
also excite these modes.  In the process, we also show how
gravitational lensing, by both density perturbations as well as GWs,
can be described in terms of even- and odd-parity BiPoSHs
(Sec.~\ref{sec:lensing}). Finally, we discuss how odd-parity BiPoSHs
could be used as probes of parity violation, and consider in
particular the cross-correlation of opposite parity CMB lensing and
polarization components (Sec.~\ref{sec:chiralGW}). We calculate the
anticipated spectra and errors for such correlations, and determine
that a large signal-to-noise is not expected for these
cross-correlations, given the current upper bounds on a GW background.

\section{Review of Bipolar Spherical Harmonics} \label{sec:review}

\subsection{Statistically Isotropic and Gaussian Maps}

A CMB temperature map $T(\hat n)$, as a function of
position $\hat n$ on the sky, can be decomposed into
spherical-harmonic coefficients
\begin{equation}
     a_{lm} = \int\, d^2 \hat n\, T(\hat n) \, Y_{lm}^*(\hat n).
\nonumber
\end{equation}
If the map is statistically isotropic and Gaussian, then the
statistics can be determined entirely in terms of the power
spectrum $C_l$, defined by
\begin{equation}
     \VEV{a_{lm} a^*_{l'm'}} = C_l \, \delta_{ll'} \, \delta_{mm'},
\label{eqn:powerspectrum}
\end{equation}
where the angle brackets denote an average over all
realizations, and $\delta_{ll'}$ and $\delta_{mm'}$ are
Kronecker deltas.  Eq.~(\ref{eqn:powerspectrum}) states that all
of the $a_{lm}$ are uncorrelated, and Gaussianity further
dictates that the probability distribution function for any $a_{lm}$ to take on a
particular value is a Gaussian distribution with variance $C_l$.

The spatial temperature autocorrelation function is defined to be
$C(\hat n,\hat n') \equiv \VEV{T(\hat n) \,T(\hat n')}$.  Most
generally it is a function of the two directions $\hat n$ and
$\hat n'$.  However, if the map is statistically isotropic and
Gaussian, then the spatial correlation function depends only on the
angle $\theta$, given by $\cos\theta=\hat n \cdot \hat n'$,
between the two directions.  In this case,
\begin{equation}
     C(\hat n,\hat n') = \sum_l \frac{(2l+1)}{4\pi} C_l
     P_l(\hat n\cdot \hat n'),
\nonumber
\end{equation}
where $P_l(x)$ are the Legendre polynomials.

\subsection{Departures from Gaussianity/SI}

Departures from Gaussianity and/or SI will
induce correlations between different $a_{lm}$s.  The most
general correlation between any two $a_{lm}$s can be written,
\begin{equation}
     \VEV{ a_{lm} a^*_{l'm'}} = C_l \delta_{ll'} \delta_{mm'}
     +\sum_{LM;L>0}  (-1)^{m'} \cleb{L}{M}{l}{m}{l',}{-m'} \ALM,
\label{bipoSHexp}
\end{equation}
where $C_l$ is the (isotropic) power spectrum,
$\cleb{L}{M}{l}{m}{l'}{m'}$ are Clebsch-Gordan coefficients, and
the $\ALM$ are BiPoSH coefficients.  The spatial two-point
correlation function is then
\be
     C(\hat n, \hat n') = \sum_l \frac{(2l+1)}{4\pi} C_l P_l(\hat n\cdot \hat n') + \sum_{l l' L M} \,\ALM \,\,\{Y_l(\hat n)\otimes Y_{l'}(\hat n')\}_{LM},
\ee
where 
\be
     \{Y_l(\hat n)\otimes Y_{l'}(\hat n')\}_{LM} = \sum_{mm'}
     \cleb{L}{M}{l}{m}{l'}{m'} \, Y_{lm}(\hat n) Y_{l'm'}(\hat
     n'),
\label{eqn:biposhs}
\ee
are the bipolar spherical harmonics (BipoSHs).  These BiPoSHs
constitute a complete orthonormal basis for functions of $\hat n$
and $\hat n'$ in terms of total-angular-momentum states labeled
by quantum numbers $L$ and $M$ composed of angular-momentum
states with $lm$ and $l'm'$; they are an alternative to the outer
product of the $\{l,m\}$ and $\{l',m'\}$ bases.

\subsection{Odd-Parity Bipolar Spherical Harmonics} \label{sec:oddbiposh}

It is instructive to decompose $\ALM$ into its odd and even parity parts,
\begin{equation}
     \ALM =  \ALMp  \frac{[1+ (-1)^{l+l'+L}]}{2} + \ALMm
     \frac{[1- (-1)^{l+l'+L}]}{2},
\label{bparity}
\end{equation}
where $\ALMp$ ($\ALMm$) are zero for the sum $l+l'+L$ being odd
(even). It follows from the symmetry $C(\hat n,\hat n')=C(\hat
n',\hat n)$ that $\ALMp$ ($\ALMm$) are (anti) symmetric in $l$
and $l'$. We also infer that $\Big[\ALMp\Big]^* = (-1)^M
\ALmMp $ and $\Big[\ALMm\Big]^* = (-1)^{M+1}\ALmMm$. Thus, odd-parity BiPoSHs vanish for $l=l'$. Prior literature
has considered physical effects (e.g., nontrivial topologies \cite{Souradeep:2006kk},
SI violation \cite{Joshi:2009mj,Aich:2010gn}) that produce only $\ALMp$, the
even-parity BiPoSHs, and measurements have been carried out with
WMAP data only for the $\ALMp$ \cite{Hajian:2004zn, Souradeep:2005cq}.  In this paper, we consider also
the odd-parity BiPoSHs $\ALMm$.

Estimators for the BiPoSH coefficients (both the $\oplus$ and $\ominus$
modes) can be constructed from a map of the CMB temperature field $T(\hat n)$, as
follows:
\begin{equation}
     \widehat{\ALM} = \sum_{mm'} W_l^{-1} W_{l'}^{-1} \: a^{\rm map}_{lm} a^{*\,{\rm map}}_{l'm'} (-1)^{m'}
     \cleb{L}{M}{l}{m}{l',}{-m'},
\label{eqn:ALMestimator}
\end{equation}
and this estimator has a variance, under the null hypothesis (a 
SI Gaussian map),
\begin{equation}
     \VEV{ \widehat{\ALM} \widehat{\ALMt{L'}{M'}{\bar{l}}{\bar{l'}}}^*} = \delta_{LL'} \: \delta_{MM'} \:\left[\delta_{l\bar{l}}\delta_{l'\bar{l'}} + (-1)^{l+l'+L}\delta_{l\bar{l'}}\delta_{\bar{l}l'}\right] \,
     C_l^{\mathrm{map}} C_{l'}^{\mathrm{map}}\,W_l^{-2}\,W_{l'}^{-2},
\label{eqn:ALMvariance}
\end{equation}
where $a_{lm}^{\mathrm{map}} = W_l \, a_{lm} + a_{lm}^{\rm noise}$ and $C_l^{\mathrm{map}} = W_l^2 \, C_l + N_l$ are the temperature spherical-harmonic coefficients and power spectrum corrected for detector noise and finite resolution. The Gaussian detector window function, which encapsulates the effects of finite detector resolution, is given by $W_l \equiv \exp \left[ -l^2 \theta_{\rm FWHM}^2/(16 \ln 2) \right]$, where $\theta_{\rm FWHM}$ is the full width at half maximum of the detector. The instrumental noise contribution to the temperature power spectrum is given by
\be N_l = \frac{4 \pi ({\rm NET})^2}{t_{\rm obs} \sqrt{f_{\rm sky}}}, \nonumber\ee 
where $f_{\rm sky}$ is the fraction of the sky observed, ${\rm
NET}$ is the noise equivalent temperature of the detector, and
$t_{\rm obs}$ is the length of time over which the CMB was
observed by a particular survey. We notice that the variance in
Eq.~(\ref{eqn:ALMvariance}) vanishes for odd parity and
$l=l'=\bar{l}=\bar{l'}$, which is expected given that odd-parity
BiPoSHs with $l=l'$ vanish. 

The noise in any individual $\ALM$ is large, and so a
search for a statistically significant departure from zero in
one or a handful of $\ALM$s will probably not be too effective.
It is better to consider specific models and/or parameterizations
for departures from SI/Gaussianity and then combine the $\ALM$s
into a minimum-variance estimator for the
SI/Gaussianity-violating parameters of those models.  For
example, Ref.~\cite{Hajian:2003,Hajian:2005jh} considered the bipolar power
spectrum $\kappa_L \equiv \sum_{ll'M} |\ALM|^2$ as a
parameterization for departures from SI.  As
another example, Ref.~\cite{Pullen:2007tu} combined $\ALM$s with
$L=2$ and $l'=l,l\pm2$ to derive minimum-variance estimators for
the amplitude of an inflation-induced primordial-power quadrupole
of the type considered in Ref.~\cite{Ackerman:2007nb}.

\section{Gravitational Lensing} \label{sec:lensing}
\subsection{Gradient and Curl-Type Deflections}\label{sec:deflecdecomp}

Consider a statistically isotropic and homogeneous Gaussian
temperature map $T_g(\hat{n})$ on the sphere, where $\hat{n}$ is a position on the sky.  Now suppose
that each point on the sky $\hat{n}$ has been deflected from an original direction $\hat{n} + \vec{\Delta}(\hat{n})$ so that the observed temperature
is $T(\hat{n})= T_g(\hat{n}+\vec{\Delta}) \simeq
T_g(\hat{n}) + \vec\Delta \cdot \delt
T_g(\hat{n})$.  This deflection might come about
cosmologically through weak gravitational lensing or may arise as
an instrumental/measurement artifact (for example, if there are
pointing errors).

The most general deflection field $\vec{\Delta}$ can be
written in vector notation as
\begin{equation}
     \vec{\Delta} = \delt \: \phi(\hat{n}) + \delt \times
     \Omega(\hat{n}),
\label{eqn:displacements}
\end{equation}
or in component notation, $\Delta_i = \left(\nabla_{\vec{\theta}}\right)_i
\phi(\hat{n}) + \epsilon_{ij} \left(\nabla_{\vec{\theta}}\right)_j \Omega(\hat{n})$,\footnote{Here, the Levi-Civita symbol on the unit sphere can be defined in terms of its three-dimensional equivalent as $\epsilon_{ij} = -\epsilon_{ijk} \, r_k$. The choice of sign here can be understood as the choice to have the spherical polar coordinates ($\theta$, $\phi$) form a right-handed coordinate system on the sky, since it will ensure that the basis vectors satisfy $\hat{\rm e}_{\theta} \times \hat{\rm e}_{\phi} = 1$.} in terms of two scalar functions $\phi(\hat{n})$ and
$\Omega(\hat{n})$ on the sphere, where $\delt$ is the angular covariant derivative on the unit sphere.  In other words, the most
general vector field on a two-sphere can be written as the
gradient of some scalar field $\phi(\hat{n})$ plus the curl of some other field $\Omega(\hat{n})$.  Weak gravitational lensing by density
perturbations gives rise, at linear  order in the lensing
potential, only to the gradient component.  A curl component can
arise cosmologically from second-order terms in the deflection field
or from lensing by GWs.  Systematic measurement
effects may conceivably give rise to both types of deflections.

We now show that the $\ALMp$ and $\ALMm$ BiPoSH coefficients are
induced, respectively, by the gradient and curl components of the deflection field.
The change in the temperature moments induced by lensing is (at
first order in $\phi$ and $\Omega$),
\begin{eqnarray}
     \delta a_{lm} &=& \int d^2\hat{n} \, Y_{lm}^*(\hat n) \left\{\left[\delt \: \phi\right]\cdot \left[ \delt \: T(\hat n) \right] + \left[\delt \: \Omega(\hat n)\right]  \times \left[\delt \: T(\hat n) \right] \right\} \nonumber \\
     & = & \!\!\!\! \sum_{LM;\,L>0} \: \sum_{l' m'}  a_{l'm'} \!\! \int d^2\hat{n} \, Y_{lm}^*(\hat n) \left\{\phi_{LM} \! \left[\delt \: Y_{LM}(\hat n) \right]\!\cdot\! \left[\delt \: Y_{l'm'}(\hat n) \right] \!+ \Omega_{LM}\!\left[ \delt \: Y_{LM}(\hat n) \right] \!\times\! \left[ \delt \: Y_{l' m'}(\hat n) \right] \right\}\!\!,
\nonumber\end{eqnarray}
where in the second line we have decomposed 

\be \phi(\hat{n}) = \sum_{L=1}^{\infty} \sum_{M=-L}^L Y_{LM}(\hat{n}) \: \phi_{LM}, \label{eq:phidecomp}\ee 

\noindent and similarly for $\Omega(\hat{n})$.  We do not consider $L=0$ modes of $\phi$ and $\Omega$ since they would not cause a deflection. In the notation of Ref.~\cite{Hu:2000ee},
\begin{equation}
     \delt \: Y_{lm} = \sqrt{\frac{l(l+1)}{2}}
     \left[{}_1Y_{lm} \, \hat m_+ - {}_{-1}Y_{lm} \, \hat m_- \right],
\nonumber\end{equation}
where ${}_1Y_{lm}$ and ${}_{-1}Y_{lm}$ are spin-weighted spherical harmonics, the null coordinates $\hat m_{\pm} = (\hat e_\theta \mp i \hat
e_\phi)/\sqrt{2}$, and the only non-trivial products of the null coordinates are $\hat m_+ \cdot \hat m_-
= 1$, and $\hat m_+ \times \hat m_- = i$.
Thus, it is obtained that
\begin{eqnarray}
     \left(\delt \: Y_{LM}\right) \cdot \left(\delt \: Y_{l' m'}\right) &= &-\frac{\sqrt{L(L+1)l'(l'+1)}}{2} \left[\left({}_1Y_{LM} \right) \left({}_{-1} Y_{l' m'} \right) + \left({}_{-1} Y_{LM}\right) \left({}_1Y_{l' m'} \right) \right], \nonumber\\
     \left(\delt \: Y_{LM}\right) \times \left(\delt \: Y_{l' m'}\right) &= & -\frac{i \sqrt{L(L+1)l'(l'+1)}}{2} \left[\left({}_1Y_{LM} \right) \left({}_{-1} Y_{l' m'} \right) - \left({}_{-1} Y_{LM}\right) \left({}_1Y_{l' m'} \right) \right].\nonumber
\end{eqnarray}
Using the triple integral  \cite{Hu:2000ee} of spin-weighted spherical
harmonics, the $\delta a_{lm}$ for the
gradient and curl terms are obtained as 
\begin{equation}
     \delta a_{lm} = \sum_{LM;\,L>0} \:  \sum_{l'm'} \frac{(-1)^{M+m} \: a_{l'm'} \, G^L_{ll'}}{\sqrt{(2L + 1)l(l+1)}} \left[ \phi_{LM}\frac{\left[1+ (-1)^{l+l'+L}\right]}{2} - i \, \Omega_{LM} \frac{\left[1- (-1)^{l+l'+L}\right]}{2}\right] \cleb{L}{M}{l}{m}{l',}{-m'},
\nonumber
\end{equation}
where
\begin{equation}
     G_{ll'}^L \equiv \sqrt{\frac{L(L+1)l(l+1)l'(l'+1)
     (2l+1)(2l'+1)}{4\pi}} \cleb{L}{1}{l}{0}{l'}{1}.
\nonumber
\end{equation}
Up to linear order in the deflection coefficients $\phi_{LM}$ and $\Omega_{LM}$, the
even- and odd-parity BiPoSH coefficients are then,
\begin{eqnarray}
     \ALMp &=& \frac{\phi_{LM}}{\sqrt{2L+1}}\,\left[\frac{C_l G^L_{l'l}}{\sqrt{l'(l'+1)}} + \frac{C_{l'} G^L_{ll'}}{\sqrt{l(l+1)}} \right] = Q^{\oplus L}_{ll'} \phi_{LM}, \label{eqn:evenparity}  \\
     \ALMm &=& \frac{i \Omega_{LM}}{\sqrt{2L+1}}\,\left[\frac{C_l G^L_{l'l}}{\sqrt{l'(l'+1)}} - \frac{C_{l'} G^L_{ll'}}{\sqrt{l(l+1)}}\right] = Q^{\ominus L}_{ll'} \Omega_{LM},\label{eqn:oddparity}
\end{eqnarray}
where we have defined the quantities
\begin{align} 
Q^{\oplus L}_{ll'} &= \frac{1}{\sqrt{2L+1}} \left[\frac{C_l G^L_{l'l}}{\sqrt{l'(l'+1)}} + \frac{C_{l'} G^L_{ll'}}{\sqrt{l(l+1)}}\right], \nonumber\\
Q^{\ominus L}_{ll'} &= \frac{i}{\sqrt{2L+1}} \left[ \frac{C_l G^L_{l'l}}{\sqrt{l'(l'+1)}} - \frac{C_{l'} G^L_{ll'}}{\sqrt{l(l+1)}}\right].
\nonumber\end{align}
Clearly, the gradient part contributes only to $\ALMp$ and the curl part only to $\ALMm$. Further, it is explicit that the gradient and curl parts of the deflection correspond, respectively, to the symmetric and antisymmetric (in $\{l l'\}$) parts of the total $\ALM$.

Suppose the $\ALM$s have been measured using the estimators in
Eq.~(\ref{eqn:ALMestimator}).  If we then assume that lensing is the dominant source of BiPoSHs we can use Eqs. (\ref{eqn:ALMvariance}), (\ref{eqn:evenparity}), and (\ref{eqn:oddparity}) to construct maximum-likelihood estimators for the gradient and curl components of the deflection field,

\be  \widehat{\phi_{LM}} = \frac{\sum_{ll'} Q^{\oplus L*}_{ll'} \widehat{\ALMp}\big/\left( W_l^{-2} W_{l'}^{-2} C^{\rm map}_l C^{\rm map}_{l'} \right)}{\sum_{ll'} \left|Q^{\oplus L}_{ll'}\right|^2/\left( W_l^{-2} W_{l'}^{-2} C^{\rm map}_l C^{\rm map}_{l'} \right)},  \label{eq:phest}\ee

\be  \widehat{\Omega_{LM}} = \frac{\sum_{ll'} Q^{\ominus L*}_{ll'} \widehat{\ALMm}\big/\left( W_l^{-2} W_{l'}^{-2} C^{\rm map}_l C^{\rm map}_{l'} \right)}{\sum_{ll'} \left|Q^{\ominus L}_{ll'}\right|^2/\left( W_l^{-2} W_{l'}^{-2} C^{\rm map}_l C^{\rm map}_{l'} \right)}\,.  \label{eq:omest}\ee

 \noindent The variance of these estimators, under the null hypothesis of no lensing, is given by
 
 \be \langle \widehat{\phi_{LM}} \: \widehat{\phi_{L'M'}}^*\rangle \equiv \delta_{LL'} \: \delta_{M M'} \: \big(\sigma^{\phi}_L\big)^2 \equiv 2 \: \delta_{LL'} \: \delta_{M M'} \left[\sum_{ll'} \left|Q^{\oplus L}_{ll'}\right|^2/\left( W_l^{-2} W_{l'}^{-2} C^{\rm map}_l C^{\rm map}_{l'} \right)\right]^{-1}, \label{eq:phivar}\ee

 \be \langle \widehat{\Omega_{LM}} \: \widehat{\Omega_{L'M'}}^*\rangle \equiv \delta_{LL'} \: \delta_{M M'} \: \left(\sigma^{\Omega}_L\right)^2 \equiv 2 \: \delta_{LL'} \: \delta_{M M'} \left[\sum_{ll'}\left| Q^{\ominus L}_{ll'}\right|^2/\left( W_l^{-2} W_{l'}^{-2} C^{\rm map}_l C^{\rm map}_{l'} \right)\right]^{-1}, \label{eq:Omvar} \ee
 
 \noindent where the sums in Eqs. (\ref{eq:phest}) and (\ref{eq:phivar}) only include pairs of $l$, $l'$ for which $l+l'+L$ is even, while those in Eqs. (\ref{eq:omest}) and (\ref{eq:Omvar}) only include pairs for which this quantity is odd.

\subsection{Deflection Field from Metric Perturbations}

Cosmic shear, weak gravitational lensing due to density perturbations or
GWs along the line of sight to the CMB, will
produce displacements like those in
Eq.~(\ref{eqn:displacements}).  Our goal here will be to
calculate the displacement spherical-harmonic coefficients
$\phi_{LM}$ and $\Omega_{LM}$ that arise from gravitational
lensing due to density perturbations and GWs.
There is a vast literature on lensing by density perturbations
and also specifically on lensing of the CMB by density
perturbations \cite{Lewis:2006fu}.  Our density-perturbation results follow
most closely those of Refs.~\cite{Stebbins:1996wx,Hu:2000ee}.
Lensing by GWs has been considered in
Ref.~\cite{Kaiser:1996wk}.  We follow primarily the approach of
Refs.~\cite{Dodelson:2003bv,Dodelson:2010qu}, who calculated
$\Omega_{LM}$ due to GWs, but extend their
results to include $\phi_{LM}$ from GWs,
reproducing the results of Ref.~\cite{Li:2006si}.  We make use
in this Section of relevant work on lensing and/or differential
analysis on the celestial sphere in
Refs.~\cite{Hu:2000ee,Stebbins:1996wx,Kamionkowski:1996ks,Kamionkowski:1996zd}.

We write the metric for the perturbed spacetime as

\be ds^2 = a^2(\eta)\left[ -d\eta^2 + \left(\delta_{ij} + h_{ij}\right) dx^i dx^j \right], \nonumber\ee

\noindent where $h_{ij}$ is the metric perturbation in the synchronous gauge, and $\eta$ is the conformal time. Now consider a photon that we observe to come from the direction $\hat{n}$ on the sky. In the absence of perturbations, this photon travels along a path $\vec{x}(\eta) = (\eta_0 - \eta) \, \hat{n}$ as a function of conformal time $\eta$, where $\eta_0$ is the conformal time today. Metric perturbations will induce perturbations in this trajectory, which we can calculate by integrating the geodesic equation back over the photon path to find the direction of propagation of the photon when it was emitted at a conformal time $\eta$. To first order in the metric perturbation $h$, we find the original direction of propagation of the photon on the sky to be $\hat{n} + \vec{\Delta}$, where \cite{Book:2010pf}

\be \Delta^i(\hat{n}) = \frac{P_{im}}{\eta_0 - \eta} \int_{\eta_0}^{\eta} d\eta' \left[ h_{mj} \hat{n}_j - \frac{1}{2} (\eta' - \eta) \hat{n}_k \hat{n}_l \partial_m h_{kl} \right]_{[\eta', \; (\eta_0 - \eta') \hat{n}]}. \label{eqn:trajectory}\ee

\noindent Here, we have ignored the observer terms $h_{ij}(\eta_0)$, and we have defined the projection tensor $P_{im} = \delta_{im} - n_i n_m$ onto the space perpendicular to the unit vector $\hat{n}$. The subscript indicates that the quantities in the integral are evaluated at time and space coordinates $(\eta, \vec{x}) =(\eta', \; (\eta_0 - \eta') \hat{n})$; i.e. they are evaluated along the unperturbed path of the photon. In our case, the source is the CMB, and $\eta = \eta_{\rm lss}$ is the conformal time at the surface of last scatter. However, the calculation could also be applied to the lensing of galaxies in which case the relevant conformal time would be that corresponding to redshifts $z\sim1$.

The functions $\phi(\hat{n})$ and $\Omega(\hat{n})$ in the
decomposition in Eq. (\ref{eqn:displacements}) can be obtained from

\be \nabla^{2}_{\vec{\theta}} \, \phi(\hat{n}) = \vec{\nabla}_{\vec{\theta}} \cdot \vec{\Delta}(\hat{n}), \quad \quad \nabla^{2}_{\vec{\theta}} \, \Omega(\hat{n}) = - \vec{\nabla}_{\vec{\theta}} \times \vec{\Delta}(\hat{n}), \label{eq:phiomfromdelt}\ee

\noindent where as before $\vec{\nabla}_{\vec{\theta}}$ is the angular covariant derivative on the unit sphere. As Ref. [25] notes, the standard lensing convergence is $\kappa = -(1/2) \nabla^2_{\vec{\theta}}\;\phi$ and the lensing rotation is $\omega = (1/2) \nabla^2_{\vec{\theta}}\;\Omega$. 

The gradient component is obtained from 

\begin{align} \nabla^{2}_{\vec{\theta}} \; \phi(\hat{n}) =& \vec{\nabla}_{\vec{\theta}} \cdot \vec{\Delta} = -\frac{1}{\eta_0 - \eta} \Bigg\{ \int_{\eta}^{\eta_0} d\eta' (\eta_0 - \eta') (\delta_{ik} - \hat{n}_i \hat{n}_k) \left[ -\partial_k \left( h_{ij} n^j \right) + \frac{1}{2} (\eta' - \eta) \partial_i \partial_k (h_{lm} \hat{n}_l \hat{n}_m) \right]_{[\eta', \; (\eta_0 - \eta') \hat{n}]} \nonumber\\ &+ \int_{\eta}^{\eta_0} d\eta' \left[ 3 \hat{n}_i \hat{n}_j h_{ij} -h_{ii} + (\eta' - \eta)  \left( \hat{n}_j \partial_i h_{ij} - 2 \: \hat{n}_i \hat{n}_j \hat{n}_k \partial_k h_{ij} \right) \right]_{[\eta', \; (\eta_0 - \eta') \hat{n}]}  \Bigg\}, \label{eqn:grad}\end{align}

\noindent where we have used the fact that $\vec{\nabla}_{\vec{\theta}}$, which acts on the unit vector $\hat{n}$, behaves as $\nabla^i_{\vec{\theta}} = (\eta_0 - \eta') (\delta_{ik} - \hat{n}_i \hat{n}_k) \partial_k$ inside the integral due to the dependence of $\vec{x}$ on $\hat{n}$ as defined in the integrand subscript.

Let us now consider the curl component.  For this calculation we must use $\nabla_{\vec\theta}^2 \Omega = -\delt\times \vec{\Delta}$ and then note that, as before, $\nabla^i_{\vec{\theta}} = (\eta_0 - \eta') (\delta_{ik} - \hat{n}_i \hat{n}_k) \partial_k$ inside the integrand. Applying this to Eq.~(\ref{eqn:trajectory}), we have \cite{Dodelson:2003bv}
\begin{equation}
     \nabla_{\vec\theta}^2 \Omega(\hat n) = - \int_{\eta}^{\eta_0} \, d\eta'\, (n_i n_l \epsilon_{ijk} \partial_j h_{kl})_{\left[\eta',\hat n(\eta_0 -\eta') \right]}.
\label{eqn:curl}
\end{equation}

\subsection{Lensing by Density (Scalar Metric) Perturbations}

Let us first consider scalar perturbations. In the conformal-Newtonian gauge in the absence of anisotropic stresses, the metric is given by

\be ds^2 = a^2(\eta)\left[ -(1 - 2 \Phi) d\eta^2 + (1 + 2 \Phi) \delta_{ij} dx^i dx^j \right]. \nonumber\ee

\noindent Noting that a conformal transformation preserves null geodesics, our calculations of the photon path will be unaffected if we work in a synchronous metric obtained from the conformal-Newtonian form through multiplication by $(1 + 2 \Phi)$. Assuming that $\Phi$ is small and keeping terms only to linear order, we find the conformally related metric,

\be  ds^2 = a^2(\eta)\left[ -d\eta^2 + (1 + 4 \Phi) \delta_{ij} dx^i dx^j \right]. \nonumber\ee

\noindent Using this metric perturbation $h_{ij} = 4 \Phi \delta_{ij}$ in Eq. (\ref{eqn:grad}) above, we find that the first, third, and fourth terms vanish, giving for the gradient-type lensing caused by scalar perturbations,

\be \nabla^{2}_{\vec{\theta}} \; \phi^{\rm sca}(\hat{n}) = - \frac{2}{\eta_0 - \eta} \int_{\eta}^{\eta_0} d\eta' (\eta' - \eta) \big[ \left(\delta_{ij} - \hat{n}_i \hat{n}_j \right) (\eta_0 - \eta') \partial_i \partial_j \Phi - 2 \, \hat{n}_i \, \partial_i \Phi \: \big]. \nonumber\ee

\noindent For small-scale fluctuations, the second term will be negligible compared with the first, so it can be dropped.  We can rewrite the spatial derivatives in terms of $\vec{\nabla}_{\vec{\theta}}$ to find

\be  \nabla^{2}_{\vec{\theta}} \; \phi^{\rm sca}(\hat{n}) = - \frac{2}{\eta_0 - \eta} \int_{\eta}^{\eta_0} d\eta' \frac{\eta' - \eta}{\eta_0 - \eta'} \nabla^{2}_{\vec{\theta}} \; \Phi\big(\eta', \; (\eta_0 - \eta') \hat{n}\big), \nonumber\ee

\noindent and we can remove the angular derivatives to obtain the usual expression for the projected potential

\be \phi^{\rm sca}(\hat{n}) = -2 \int_{\eta}^{\eta_0} d\eta' \frac{\eta' - \eta}{(\eta_0 - \eta) (\eta_0 - \eta')} \Phi\big(\eta', \; (\eta_0 - \eta') \hat{n}\big). \nonumber\ee

We can once again decompose $\phi(\hat{n})$ in terms of its spherical-harmonic coefficients as in Eq.~(\ref{eq:phidecomp}).  We then find

\begin{align} \phi_{LM}^{\rm sca} &\equiv \int d^2 \hat{n} \: Y^*_{LM}(\hat{n}) \phi^{\rm sca}(\hat{n}) \nonumber\\
&=-2 \int_{\eta}^{\eta_0} d\eta' \frac{\eta' - \eta}{(\eta_0 - \eta) (\eta_0 - \eta')} \int d^2 \hat{n} \: Y^*_{LM}(\hat{n}) \: \Phi\big(\eta', (\eta_0 - \eta') \hat{n}\big). \label{eq:philmsca}\end{align}

\noindent Thus, lensing by density perturbations with a given
projected potential is characterized by nonzero even bipolar
spherical harmonics $\ALMp$ given by Eq.~(\ref{eqn:evenparity})
with $\phi_{LM}$ given by $\phi_{LM}^{\rm sca}$ above.  Scalar
perturbations cause no curl-type lensing, which we can see in
several ways. For scalar perturbations, $h_{ij} \propto \Phi \:
\delta_{ij}$, and so the left-hand side of Eq.~(\ref{eqn:curl})
vanishes. Then, by taking a Laplacian of the mode expansion
$\Omega_{LM} = \int\, d^2\hat n \, \Omega(\hat n) Y_{LM}^*(\hat
n)$, and noting that the spherical harmonics are eigenfunctions
of the Laplacian with eigenvalue $L(L+1)$, we can write

\be \Omega_{LM} = \frac{1}{L(L+1)} \int d^2{\hat n} Y^*_{LM}(\hat n) \nabla^{2}_{\vec{\theta}} \Omega(\hat{n}). \label{eq:Omlm} \ee

\noindent Thus, we find that all of the $\Omega^{\rm sca}_{LM}$,
except possibly for the unphysical $L=0$ mode,
vanish. Equivalently, an argument can be made that scalar
perturbations have no preferred direction, and so could not
generate curl-modes, which do have a preferred direction. Thus,
scalar modes produce no odd bipolar spherical harmonics
$\ALMm$.

We can go on to find the autocorrelation power spectrum of the $\phi_{LM}^{\rm sca}$. Starting from Eq.~(\ref{eq:philmsca}), we use the fact that the potential perturbations $\Phi(\eta,\vec{k})$ today are related to their primordial values $\Phi_P(\vec{k})$ by

\be \Phi(\eta,\vec{k}) = \frac{9}{10} \Phi_P(\vec{k}) \: T^{\,\rm sca}(k) \: \frac{D_1(\eta)}{a(\eta)}, \nonumber\ee

\noindent where $a(\eta)$ is the scale factor, $T^{\,\rm sca}(k)$ is the scalar transfer function that describes the evolution of scalar modes through the epochs of horizon crossing and matter-radiation equality, and $D_1(\eta)$ is the growth function that captures the scale-independent evolution of scalar modes at later times  \cite{Dodelson:2003ft}. The transfer function can be approximated using the fitting form of Ref.~\cite{Bardeen:1985tr},

\be T^{\,\rm sca}(x\equiv k/k_{\rm eq}) = \frac{\ln(1+0.17\,x)}{0.171\,x} \left[ 1 + 0.284\,x + (1.18\,x)^2 + (0.399\,x)^3 + (0.490\,x)^4 \right]^{-0.25}, \nonumber\ee

\noindent where $k_{\rm eq}$ is the wavenumber of the mode that crossed the horizon at matter-radiation equality, defined as $k_{\rm eq}~\equiv~a_{\rm eq}~H(a_{\rm eq})=\sqrt{2}~H_0~a_{\rm eq}^{-1/2}$. We can write the growth function, under the assumption of cosmological-constant dark energy, as

\be D_1(\eta) = \frac{5 \, \Omega_m}{2} \frac{H(\eta)}{H_0} \int_o^{a(\eta)} \frac{da'}{\left( a' \, H\!\left(a'\right)/H_0 \right)^3}. \nonumber\ee

\noindent We also write the autocorrelation of the primordial scalar fluctuations $\langle \Phi_P(\vec{k}) \: \Phi^*_P(\vec{k}')\rangle = (2\pi)^3 \, \delta^3\!\big(\vec{k} - \vec{k}'\big) P_{\Phi}(k)$, where the primordial power spectrum is given by

\be P_{\Phi}(k) = \frac{50 \, \pi^2}{9 \, k^3} \left( \frac{k}{H_0} \right)^{n_s - 1} \Delta_R^2 \, \left( \frac{\Omega_m}{D_1(a=1)} \right)^2. \nonumber\ee

With these ingredients, and after using the partial-wave decomposition,

\be e^{ik(\eta_0-\eta') \cos\theta} = \sum_{L=0}^{\infty} i^L (2L+1) j_L\left(k(\eta_0-\eta')\right) P_L(\cos\theta), \label{eq:partialwave}\ee

\noindent we find the autocorrelation power spectrum to be

\be C_L^{\phi\phi\,\rm sca} = \frac{2}{\pi} \left[ \frac{9}{5\,(\eta - \eta_0)} \right]^2 \int dk \, k^2 \, P_{\Phi}(k) \: T^{\,\rm sca}(k)^2 \left\{ \int_{\eta}^{\eta_0} d\eta' \, \frac{(\eta' - \eta)}{(\eta_0-\eta')} \frac{D_1\!\left(\eta'\right)}{a\!\left(\eta'\right)} j_L\left[ \left(\eta_0 - \eta'\right) k \right] \right\}^2. \ee

To calculate the magnitude and shape of this autocorrelation function, we employ the WMAP 7-year cosmological parameters of Ref.~\cite{Komatsu:2010fb}. We plot the result of our calculation in green squares in Fig.~\ref{fig:phi}.

\subsection{Lensing by GWs (Tensor Metric Perturbations)} \label{sec:GWlensing}

\begin{figure}[t!] 
\centering
\includegraphics[width=0.8\textwidth]{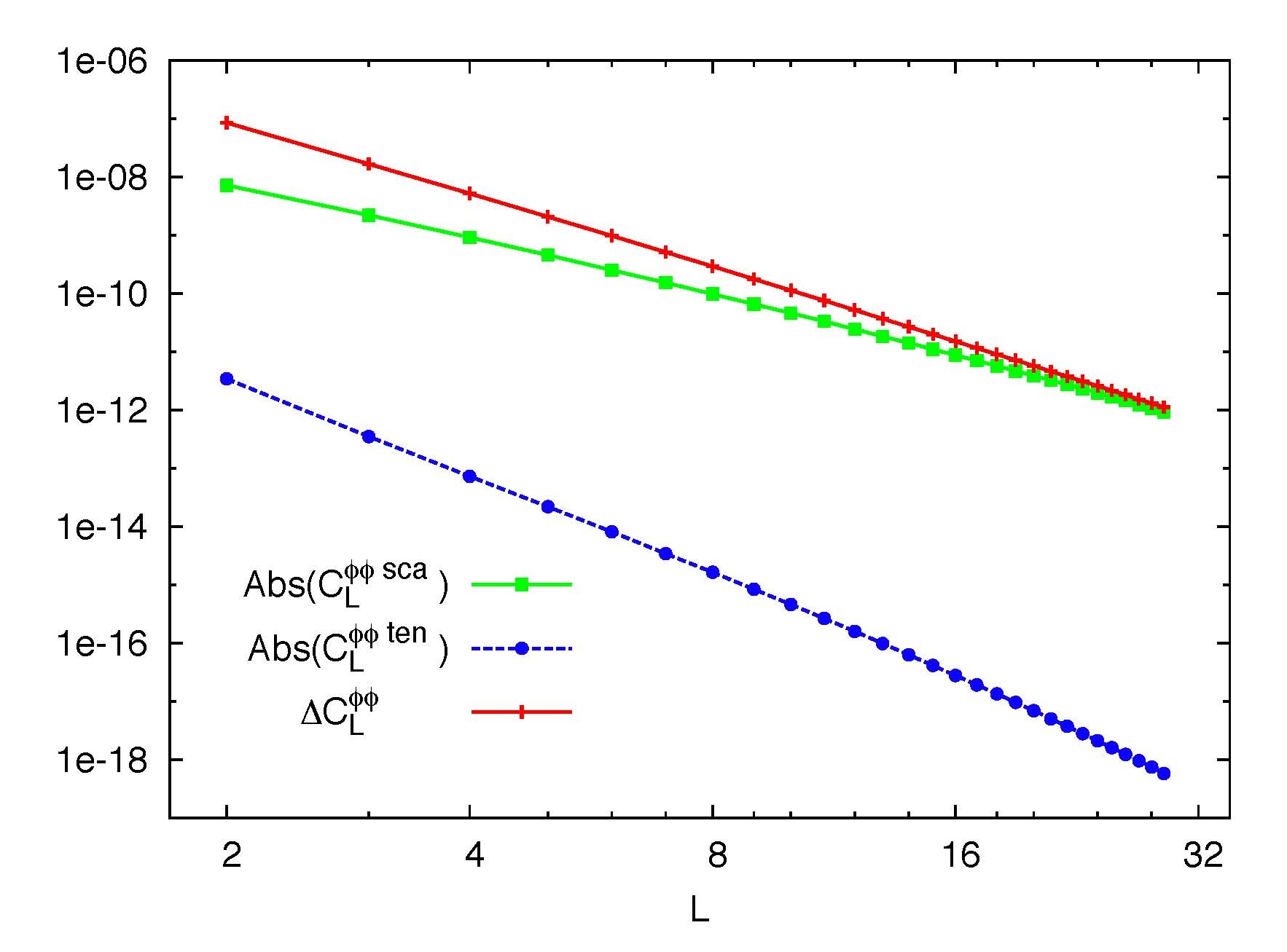}
\caption{Here we plot the autocorrelation power spectrum $C_L^{\phi\phi}$ of the gradient-type $\phi$ modes of cosmic shear. In green squares we show the autocorrelation of the $\phi$ modes from lensing by scalar perturbations, and in blue circles that of the $\phi$ modes induced by tensor perturbations. We use the WMAP-7 cosmological parameters, and assume the maximum allowable tensor-to-scalar ratio $r=0.24$ from the WMAP-7 data combined with BAO and the $H_0$ measurement \cite{Komatsu:2010fb}, to calculate the tensor contribution. The error with which these power spectra could be measured using the parameters of the Planck satellite is shown as red $+$s.}
\label{fig:phi}
\end{figure}

\begin{figure}[t!]
\centering
\includegraphics[width=0.8\textwidth]{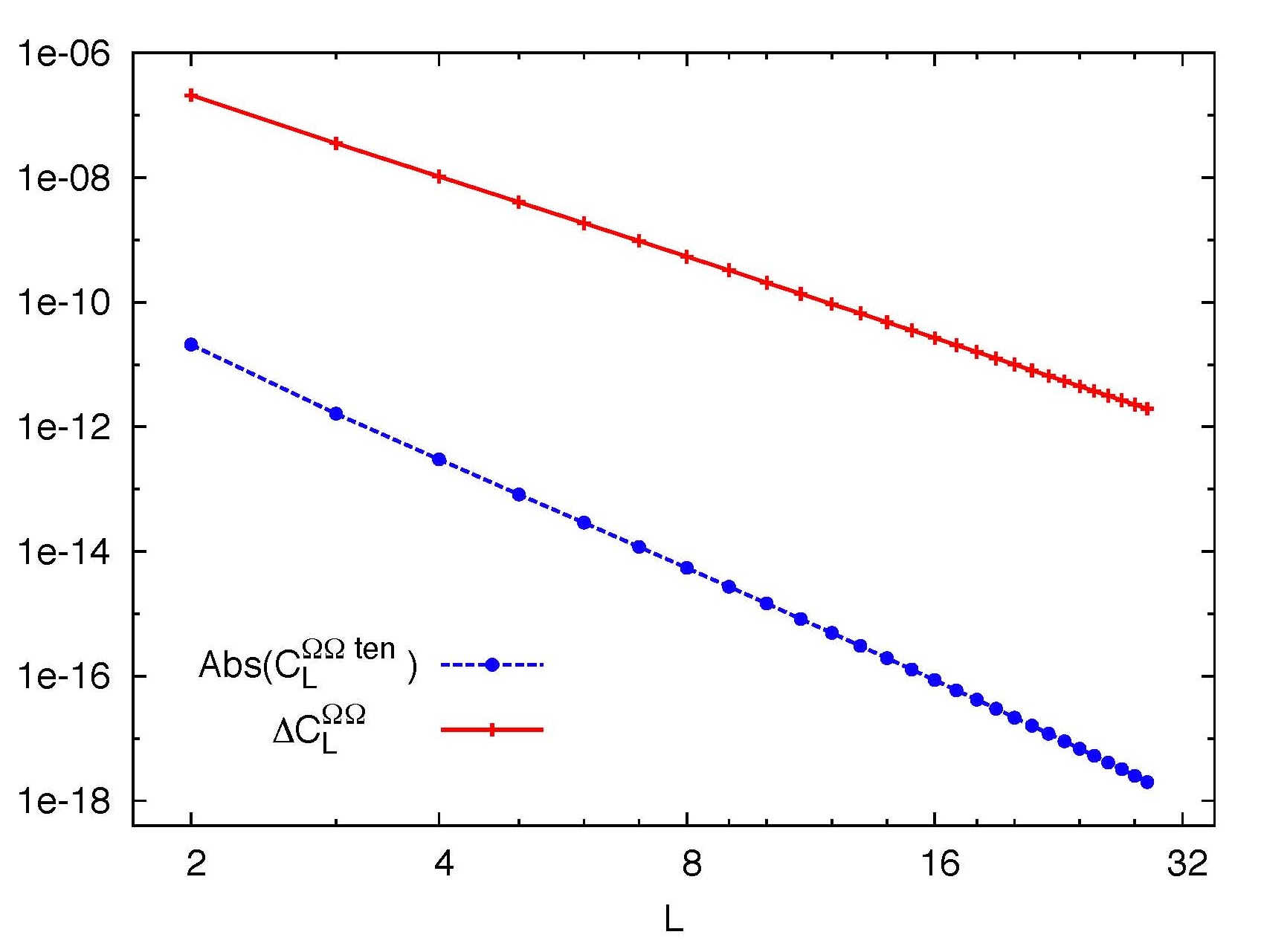}
\caption{Here we plot the autocorrelation power spectrum $C_L^{\Omega\Omega}$ of the curl-type $\Omega$ modes of the weak lensing of the CMB temperature field. These modes can only be induced by tensor perturbations. We show the signal in blue circles and the error with which they could be measured using the parameters of the Planck satellite as red $+$s.}
\label{fig:Omega}
\end{figure}

If the metric perturbation $h_{ij}$ is caused by GWs, we can decompose it into plane waves,

\be h_{ij}(\vec{x},\eta) = \int \frac{d^3 k}{(2\pi)^3} \: e^{i \vec{k}\cdot \vec{x}} \: T(k,\eta) \sum_{\alpha = +,\times} h^{\alpha}\!(\vec{k}) \: \epsilon^{\alpha}_{ij}(\vec{k}), \label{eq:hexp} \ee

\noindent where we sum over the two  GW polarizations $+$ and $\times$, the plane-wave amplitudes are $h^{\alpha}\!(\vec{k})$, and $\epsilon^{\alpha}_{ij}$ are the polarization tensors, which are transverse, traceless matrices. Here, $T(k,\eta)$ is the GW transfer function, which gives the conformal-time evolution of the mode; Ref.~\cite{Dodelson:2003bv} notes that it is well approximated by $T(k,\eta) = 3 j_1(k\eta)/(k\eta)$.

Now consider a single GW mode
propagating in the $\hat z$ direction with wavenumber $k$ and
$+$ polarization. In this case the polarization tensor is

\be \epsilon^+_{ij}(k \hat{z}) = \begin{pmatrix} 1 & 0 & 0 \\ 0 & -1 & 0 \\ 0 & 0 & 0 \end{pmatrix}. \nonumber \ee

\noindent The only non-zero metric-perturbation
components are then $h_{xx}=-h_{yy} =  h^+\!({\vec k}) \, e^{ikz} \, T(k,\eta)$. The unit vector $\hat n
= (\sin\theta \cos\varphi,\sin\theta \sin\varphi,\cos\theta)$.  
The curl component of lensing of the CMB by tensor perturbations is then
\begin{equation}
     \nabla_{\vec\theta}^2 \: \Omega^{\rm ten}(\hat n) = i k h^+\!({\vec k}) \sin^2\theta
     \sin 2\varphi \int_{\eta}^{\eta_0} \, d\eta'\,
     T(k,\eta') e^{i
     k (\eta_0-\eta') \cos\theta}.
\label{eqn:curlcomponent}
\end{equation}
A GW with the $\times$ polarization is the same
as that with the $+$ polarization, but rotated by 45$^\circ$ to
the right.  The $\Omega^{\rm ten}(\hat n)$ pattern is therefore the same,
but with $\sin2\varphi$ replaced by $-\cos2\varphi$.  We thus
see that {\it lensing by GWs will give rise to
nonvanishing $\ALMm$}.

The gradient component of cosmic shear due to tensor perturbations is a bit more complicated; it is
\begin{align} \nabla_{\vec\theta}^2 \: \phi^{\rm ten}(\hat n) = - \frac{h^+\!({\vec k})}{\eta_0 -\eta} \sin^2\theta \cos2\varphi \int_{\eta}^{\eta_0}\, d\eta'\, T(k,\eta') \bigg\{ 3 &- 2 i k (\eta'-\eta) \cos\theta \nonumber\\
&+ (\eta_0-\eta') \left[ ik \cos\theta - \frac{k^2}{2} (\eta'-\eta) \sin^2\theta \right] \bigg\} \: e^{i k (\eta_0-\eta') \cos\theta}. \label{eqn:gradientcomponent} \end{align}
This can be further simplified by noting that 
\begin{equation}
     -ik\cos\theta e^{i k (\eta_0-\eta') \cos\theta} =
     \frac{\partial}{\partial \eta'} e^{i k (\eta_0-\eta')
     \cos\theta},
\nonumber
\end{equation}
which then leads to
\begin{align} \nabla_{\vec\theta}^2 \: \phi^{\rm ten}(\hat n) = - \frac{h^+\!({\vec k})}{\eta_0 -\eta} \sin^2\theta \cos2\varphi \int_{\eta}^{\eta_0}\, d\eta'\, T(k,\eta') \bigg\{ 3 &+ 2 \, (\eta'-\eta) \frac{\partial}{\partial \eta'} \nonumber\\
&- (\eta_0-\eta') \left[ \frac{\partial}{\partial \eta'} + \frac{(\eta'-\eta)}{2} \left( k^2 + \frac{\partial^2}{\partial \eta'^2} \right) \right] \bigg\} \: e^{i k (\eta_0-\eta') \cos\theta}. \label{eqn:gradientcomponentone} \end{align}
For the $\times$ polarization, we replace $\cos2\varphi$ by
$\sin2\varphi$.

Note that the expressions for $\nabla_{\vec\theta}^2 \: \phi^{\rm ten}$ and
$\nabla_{\vec\theta}^2 \: \Omega^{\rm ten}$ differ only in two ways: (1) The
curl mode has a $\sin2\varphi$ dependence on the azimuthal
angle $\varphi$, while the scalar mode has a $\cos2\varphi$
dependence (for the $+$ polarization).  (2) The $\eta'$
dependences of the two integrands differ.

We now find the spherical-harmonic coefficients
$\phi^{\rm ten}_{LM} = \int\, d^2\hat n \, \phi^{\rm ten}(\hat n) Y_{LM}^*(\hat n)$
and $\Omega^{\rm ten}_{LM} = \int\, d^2\hat n \, \Omega^{\rm ten}(\hat n)
Y_{LM}^*(\hat n)$.  Taking the angular derivatives of this decomposition of the curl component, we find the result Eq.~(\ref{eq:Omlm}). We also expand these coefficients in terms of their polarization and $\vec{k}$ modes,

\be \Omega^{\rm ten}_{LM} = \int \frac{d^3 k}{(2\pi)^3} \sum_{\alpha=+,\times} \Omega^{\rm ten\,\alpha}_{LM}(\vec{k}). \label{eq:Ommodeexp} \ee

\noindent If we consider just one mode, with $\alpha=+$ and $\vec{k} = k \hat{z}$, and use Eq. (\ref{eqn:curlcomponent}), its amplitude simplifies into an angular and a conformal time integral:

\be \Omega^{\rm ten \, +}_{LM}(k \hat z) = -\frac{ik\,
h^+\!({\vec k})}{L(L+1)} \int_{\eta}^{\eta_0} d\eta' \:
\: T(k,\eta') \int d^2{\hat
n} \: Y^*_{LM}(\hat n) \sin^2\!\theta \, \sin(2\phi) \:
\text{e}^{ik(\eta_0-\eta')\cos\theta}. \nonumber\ee

\noindent The azimuthal integral is easily taken once the spherical harmonic is decomposed as

\be Y^*_{LM}(\hat n) = \sqrt{\frac{2L+1}{4 \pi} \frac{(L-M)!}{(L+M)!}} \: e^{-iM\phi} \: P_{LM}(\cos\theta), \nonumber\ee

\noindent and yields the result that only $M=\pm2$ modes remain. The polar integral can then be taken by using the partial-wave decomposition Eq.~(\ref{eq:partialwave}) and by converting associated Legendre polynomials into regular Legendre polynomials and using their orthogonality. The final result that we obtain for the spherical-harmonic coefficients of the curl mode is

\be \Omega_{LM}^{\rm ten \, +}(k \hat z) = i^{L} h^+\!({\vec k}) \: (\delta_{M,2}-\delta_{M,-2}) \sqrt{\frac{2L+1}{2}} \: F_L^{\Omega}(k), \label{eqn:Omegaresult} \ee
where
\begin{equation}
     F_L^{\Omega}(k) = \sqrt{ \frac{2 \pi (L+2)!}{(L-2)!}}  \frac{1}{L(L+1)} \int_{k\eta}^{k\eta_0} \, dw\, T(w) \frac{ j_L(k\eta_0-w)} {(k\eta_0-w)^2}
\label{eqn:Omegatransfer}
\end{equation}
is a transfer function for $\Omega$.  Note that in writing
Eq.~(\ref{eqn:Omegatransfer}) we have assumed that
$T(k,\eta)=T(k\eta)$, and that for the $\times$ polarization the $\sin2\varphi$
dependence of $\Omega(\hat n)$ is replaced by $-\cos2\varphi$,
so that the factor $(\delta_{M,2}-\delta_{M,-2})$ is replaced by
$-i (\delta_{M,2}+\delta_{M,-2})$.  

Likewise, noting the similarities between Eqs.~(\ref{eqn:curlcomponent}) and (\ref{eqn:gradientcomponentone}), and decomposing $\phi^{\rm ten}_{LM}$ into modes as in Eq.~(\ref{eq:Ommodeexp})

\be \phi^{\rm ten}_{LM} = \int \frac{d^3 k}{(2\pi)^3} \sum_{\alpha=+,\times} \phi^{\rm ten\,\alpha}_{LM}(\vec{k}), \label{eq:phimodeexp} \ee

\noindent the result for the amplitude of the gradient mode with $\alpha=+$ and $\vec{k} = k \hat{z}$ is
\begin{equation}
     \phi^{\rm ten \, +}_{LM}(k \hat z) = i^{L} h^+\!({\vec k}) (\delta_{M,2}+\delta_{M,-2}) \sqrt{\frac{2L+1}{2}} F_L^{\phi}(k),
\label{eqn:phiresult}
\end{equation}
where
\begin{equation}
     F_L^{\phi}(k) = - \sqrt{ \frac{2 \pi (L+2)!}{(L-2)!}}  \frac{1}{L(L+1)} 
     \int_{k\eta}^{k\eta_0} \, dw \,
     \frac{k\eta_0-w}{k(\eta_0-\eta)} T(w) \left[
     \frac{\partial}{\partial w} + \frac{1}{2} (w-k\eta) \left(1
     + \frac{\partial^2}{\partial w^2}  \right) \right]
     \frac{ j_L(k\eta_0-w)}{(k\eta_0-w)^2}.
\label{eqn:phitransfer}
\end{equation}
Again, the factor $(\delta_{M,2}+\delta_{M,-2})$ is replaced by
$-i(\delta_{M,2}-\delta_{M,-2})$ for the $\times$ polarization.

The contributions from this Fourier mode to the $\phi^{\rm ten}$ and
$\Omega^{\rm ten}$ power spectra are $C_L^{\phi\phi \, \rm ten}(k\hat z)_+ = \sum_M
\left<|\phi^{\rm ten}_{LM}|^2\right>/(2L+1)$ and $C_L^{\Omega\Omega \, \rm ten}(k\hat z)_+ = \sum_M
\left<|\Omega^{\rm ten}_{LM}|^2\right>/(2L+1)$.  Note that it is only the $M=\pm2$ modes
that contribute.  By rotational invariance, the contribution from
the $\times$ polarization is the same, as is the contribution
from any other mode with the same wavenumber
$k$ but pointing in a different direction.  If the gravitational
waves have power spectrum $P_T(k)$, defined by
\begin{equation}
     \VEV{h_{\vec k}^i (h_{\vec k'}^j)^*} = (2\pi)^3
     \delta_D(\vec k -\vec k') \delta_{ij} P_T(k),
\label{eqn:GWpowerspectrum}
\end{equation}
(with $\{i,j\}=\{\times,+\}$), then the $\phi$ and $\Omega$ power
spectra are
\begin{equation}
     C_L^{XX \, \rm ten} = 2\int \, \frac{d^3k}{(2\pi)^3} P_T(k)
     \left[F_L^X(k)\right]^2
\label{eqn:phiOmegapowerspectra}
\end{equation}
for $X=\{\phi,\Omega\}$. In this paper, we will assume a scale-invariant power spectrum

\be P_T(k) = \frac{\pi^2}{2 k^3} \, \Delta_R^2 \, r, \ee

\noindent where we have neglected the spectral tilt and adopt the parameters of WMAP7 \cite{Komatsu:2010fb}.

We calculate the variance in the measurement of these autocorrelation functions from an observed CMB temperature map, under the null hypothesis of no GWs, and obtain an expression in terms of the variance of the $\phi$ and $\Omega$ estimators, Eqs.~(\ref{eq:phivar}) and (\ref{eq:Omvar})

\be \Delta C_L^{\phi\phi} = \sqrt{\frac{2}{2L+1}} \left(\sigma^{\phi \, 2}_L + C_L^{\phi\phi \, \rm sca}\right), \label{eq:phierr}\ee

\be \Delta C_L^{\Omega\Omega} = \sqrt{\frac{2}{2L+1}} \sigma^{\Omega \, 2}_L. \label{eq:Omegaerr}\ee

\noindent Here, $\sigma^{\phi \, 2}_L$ and $\sigma^{\Omega \, 2}_L$ are the variances of our estimators for $\phi_{LM}$ and $\Omega_{LM}$ as found in Eqs.~(\ref{eq:phivar}) and (\ref{eq:Omvar}). Note that under the null hypothesis of no GWs, there is no expected cosmological curl-type lensing signal, so this term is absent in Eq.~(\ref{eq:Omegaerr}).

To calculate these autocorrelation functions and their variances, we use the WMAP 7-year cosmological parameters of Ref.~\cite{Komatsu:2010fb}. We plot the results of our calculations in Figs.~\ref{fig:phi} and \ref{fig:Omega}, where lensing from scalar perturbations is plotted in green squares (absent for $C_L^{\Omega\Omega}$ since there are no scalar contributions to the curl modes), lensing from tensor perturbations is plotted in blue circles, and the variance of these measurements is shown in red $+$s. We use the parameters of the Planck satellite, $NET = 62 \: \mu\,K \: s^{1/2}$, $t_{\rm obs} = 2 \: {\rm yr}$, $\theta_{\rm FWHM} = 2.0635 \times 10^{-3}\: {\rm rad}$, and $f_{\rm sky} \approx 1$. We can see that the scalar $\phi$ signal is several orders of magnitude greater than the tensor signal, and that the variance with which the $\phi$-$\phi$ power spectrum could be measured with Planck is higher than the scalar signal for low multipoles. The corresponding variance with which the $\Omega$-$\Omega$ power spectrum could be measured is also significantly larger than the signal. In both of these cases, therefore, the signal to noise of measuring the lensing from tensors using all multipoles with Planck is negligible, and remains negligible even in the case of the ideal CMB experiment with zero noise.

Thus, a stochastic background of
GWs with power spectrum $P_T(k)$ predicts a
spectrum of nonzero even and odd BiPoSHs given by
Eqs.~(\ref{eqn:evenparity}) and (\ref{eqn:oddparity}), with
values of $\phi_{LM}$ and $\Omega_{LM}$ selected from Gaussian
distributions with the variances $C_L^{\phi\phi \, \rm ten}$ and
$C_L^{\Omega\Omega \, \rm ten}$ given by Eq.~(\ref{eqn:phiOmegapowerspectra}).

\subsection{BiPoSHs from Pointing Errors}

A telescope pointing error can be described as a process that causes the positions of points on the sky to be mislabeled. This then causes an effective deflection of the points on the sky ${\hat n}_{\rm obs} = {\hat n} + \vec{\Delta}\!\left({\hat n}\right)$, where ${\hat n}_{\rm obs}$ is the direction that the telescope believes it is pointed in and $\hat n$ is its actual pointing direction. As we saw in Sec.~\ref{sec:deflecdecomp}, we can decompose this deflection field $\vec{\Delta}\!\left({\hat n}\right)$ into gradient and curl components, which source even- and odd-parity BiPoSHs, respectively. Thus, from Eq.~(\ref{eq:phiomfromdelt}) we can see that any pointing error that has a nonzero curl component $\vec{\nabla}_{\vec{\theta}} \times \vec{\Delta}(\hat{n})$ will excite odd-parity BiPoSHs.

Imagine, for example, that a satellite such as Planck misestimates the rate with which it is precessing. Since it is this precession that builds up observations of subsequent rings of the sky, such a misestimation would cause a shearing of each ring relative to its neighbors. This type of a deflection has a nonzero curl component, and thus would excite odd-parity BiPoSHs. Measurement of these BiPoSHs, and in particular the odd-parity BiPoSHs, can therefore provide a useful check for such pointing errors.

\section{BiPoSHs as Probes of Parity Violation} \label{sec:chiralGW}

\subsection{Correlation of Opposite-Parity Lensing Components} \label{sec:parity}

Since the $\ALMp$ and $\ALMm$ have opposite parity for the same
$L$ and $M$, a cross-correlation between the two can arise only
if there is some parity-breaking in the physics responsible for
producing the departures from SI/Gaussianity.  Here we
mention, by way of example, chiral GWs as a
mechanism to produce such a parity-violating correlation \cite{Lue:1998mq, Contaldi:2008yz, Takahashi:2009wc}.

The contribution to the cross-correlation power spectrum from a
single Fourier mode in the $\hat z$ direction with +
polarization is $C_L^{\phi\Omega}(k\hat z) = \sum_m
\left<\phi_{LM} \: \Omega^*_{LM}\right>/(2L+1) = 0$; it vanishes as the
contribution from $M=2$ is canceled by that from $M=-2$.  And if
this is true, then by rotational invariance it is true for any
other linearly-polarized GW.  We thus conclude
that a stochastic GW background predicts
$C_L^{\phi\Omega}=0$.  In other words, there is no
cross-correlation between $\phi$ and $\Omega$, and thus no
cross-correlation between the even and odd BiPoSHs, $\ALMp$ and
$\ALMm$.

Following Ref.~\cite{Lue:1998mq}, however, consider a
right-circularly polarized GW: $h_{R} = h_+ + i
h_\times$ (i.e., we sum a $+$ polarization wave with a $\times$
polarization wave out of phase by 90$^\circ$).  The
azimuthal-angle dependence for the wave is then $e^{2i\varphi}$,
and $\Omega_{LM}$ and $\phi_{LM}$ have contributions only from
$M=2$.  There is thus a nonzero cross-correlation between $\phi$
and $\Omega$.  Similarly for a left-circularly polarized
GW $h_L=h_+ -i h_\times$, the $\varphi$
dependence is $e^{-2i\varphi}$, and only $M=-2$ modes are
excited.  There is again a cross-correlation between $\phi$ and
$\Omega$, but this time with the opposite sign. 

In the standard inflationary scenario, there are equal numbers of
right- and left-circularly polarized GWs, and
the cross-correlation between $\phi$ and $\Omega$ therefore
vanishes.  But if for some reason there is an asymmetry between
the number of right- and left-circularly polarized GWs \cite{Lue:1998mq,Contaldi:2008yz,Takahashi:2009wc,Gluscevic:2010vv},
a manifestation of parity breaking, then there may be a parity-violating
cross-correlation between $\phi$ and $\Omega$, and thus between
$\ALMp$ and $\ALMm$.  

The chirality of the GW background can be parametrized by an
amplitude $A$ which can take values between $-1$ and $1$, where
$A=+1$ denotes that all of the GWs are right-circularly
polarized, and $A=-1$ denotes that they are all left-circularly
polarized. But we have seen that a right-handed GW contributes
only to $M=2$ modes, while a left-handed one contributes only to
$M=-2$. We can denote this by weighting $M=2$ components by
$(A+1)/2$ and $M=-2$ components by $(A-1)/2$, so that our
version of Eq.~(\ref{eqn:Omegaresult}), for example, that is
appropriate to the case of a chiral GW background will be 

\be \Omega_{LM}^{\rm ten \, +}(k \hat z) = i^{L} h^+ ({\vec k}) \left[\frac{(1+A)}{2} \, \delta_{M,2} - \frac{(1-A)}{2} \, \delta_{M,-2}\right] \sqrt{\frac{2L+1}{2}} F_L^{\Omega}(k), \label{eq:OmegaAadj}\ee

\noindent and similarly for Eq.~(\ref{eqn:phiresult}). In this
way a fully right circularly-polarized GW background will have
only contributions from $M=2$, a fully left-circularly polarized
background will have only contributions from $M=-2$, and if the
amount of left and right-circularly polarized waves is equal,
that is if the GW background is non-chiral, the contributions
from $M=2$ and $M=-2$ cancel. The $\phi$-$\Omega$
cross-correlation power spectrum is given by

\be C_L^{\phi\Omega} = A \int \, \frac{d^3k}{(2\pi)^3} P_T(k) F_L^\phi(k) F_L^\Omega(k).
\label{eqn:crosspowerspectrum}
\end{equation}

Refs.~\cite{Dodelson:2003bv,Li:2006si,Cooray:2005hm} have shown that the
amplitude of the stochastic gravitational-wave background is
probably too small, even with the most optimistic assumptions,
to produce a detectable gravitational-lensing signal in the
CMB.  The example of a chiral gravitational-wave background as a
possible source of a detectable parity-breaking BiPoSH
correlation is principally of academic interest.  Still, Ref.~\cite{Dodelson:2010qu} has recently
argued that weak lensing of the CMB by GWs may
be detectable in its cross-correlation with the CMB-polarization
pattern induced by these GWs
\cite{Kamionkowski:1996ks,Kamionkowski:1996zd,Zaldarriaga:1996xe,Cabella:2004mk,Pritchard:2004qp}.
We thus surmise that a chiral gravitational-wave
background may still be able produce a detectable
parity-breaking signal in BiPoSHs in cross-correlation with the
CMB polarization, an idea we explore in the next section.

\subsection{Large-Angle CMB Polarization Spectra}

We follow the work of Ref.~\cite{Dodelson:2010qu}, finding the multipole moments of the CMB E- and B-type polarization spectra for large angular scales by considering only those modes that are produced after reionization. The spherical-harmonic coefficients of B-type polarization modes can be decomposed as

\be B_{lm} = \int \frac{d^3 k}{(2\pi)^3} \sum_{\alpha=+,\times} B_{lm}^{\alpha}\!(\vec{k}), \label{eq:Bmodeexp}\ee

\noindent where $B_{lm}^{\alpha}\!(\vec{k})$ is the amplitude of polarization B modes multipole moment $lm$ in the direction $\vec{k}$. The general form of this amplitude is quite complicated, but we can simplify it if, as in Sec.~\ref{sec:GWlensing}, we consider only a single, +-polarized GW traveling in the $\hat{z}$ direction with wavenumber $k$. In this case, the B-mode amplitude can be written

\begin{align} B_{lm}^+\!(k \hat{z}) &= i^l \: h^{\alpha}\!(\vec{k}) \: (\delta_{m,2} - \delta_{m,-2}) \: \sqrt{\frac{2l+1}{2}} \: F_l^B(k),\label{eq:Bplkz}\\
F_l^B(k) &= \frac{1}{2l+1} \: \sqrt{\frac{9\pi}{2}} \int_{\eta_{\rm re}}^{\eta_0} \! d\eta \: \dot{\tau}(\eta) \left\{ (l+2) j_{l-1}[k(\eta_0-\eta)] - (l-1) j_{l+1}[k(\eta_0-\eta)] \right\} \int_{k \eta_{\rm lss}}^{k \eta} dx \frac{-3 \, j_2(x)}{x} \frac{j_2(k \eta - x)}{(k \eta - x)^2}, \label{eq:FB} \end{align}

\noindent where the $h^{\alpha}\!(\vec{k})$ are the amplitudes of GW modes as defined in Eq.~(\ref{eq:hexp}), $\dot{\tau}(\eta)$ is the scattering rate $\dot{\tau}(\eta)~=~n_e(\eta)~\sigma_T~a(\eta)$, with $n_e$ the electron density, $\sigma_T$ the Thompson scattering cross-section, and $a$ the scale factor, and $\eta_{\rm re}$ and $\eta_0$ are the conformal times at reionization and today, respectively. Since we are only interested in small scales, we find the approximation $\eta_{\rm lss} = 0$ is sufficient for our purposes, making the last integral significantly faster to evaluate. The result above agrees with the results of Ref.~\cite{Dodelson:2010qu}, whose method we followed in its derivation, up to a factor of $i$.

We find that the corresponding E-type polarization multipoles from tensor perturbations take the same form as $B_{lm}$ above, except for the opposite sign in front of $\delta_{m,-2}$ and a different factor in the curly brackets in Eq.~(\ref{eq:FB}).  From Ref.~\cite{Pritchard:2004qp} we find this alternative form to be $(2l+1)/2 \left\{ -j_l(x) + j_l''(x) + 2 j_l(x)/x^2 + 4 j_l'(x)/x \right\}$, where here $x=[k(\eta_0-\eta)]$, and derivatives are with respect to $x$. Employing spherical Bessel function identities, we can then write the E-type polarization multipoles as

\begin{align} E_{lm} =& \int \frac{d^3 k}{(2\pi)^3} \sum_{\alpha=+,\times} E_{lm}^{\alpha}\!(\vec{k}), \label{eq:Emodeexp}\\
E_{lm}^+\!(k \hat{z}) =& i^l \: h^{\alpha}\!(\vec{k}) \: (\delta_{m,2} + \delta_{m,-2}) \: \sqrt{\frac{2l+1}{2}} \: F_l^E(k), \label{eq:Eplkz}\\
F_l^E(k) =& \frac{1}{2l+1} \: \sqrt{\frac{9\pi}{2}} \int_{\eta_{\rm re}}^{\eta_0} \! d\eta \: \dot{\tau}(\eta) \bigg\{ \frac{(2l+1)}{[k(\eta_0-\eta)]^2} j_l[k(\eta_0-\eta)] - \frac{(2l+1)(3l^2+3l-4)}{(2l-1)(2l+3)} j_{l}[k(\eta_0-\eta)] \nonumber\\
&+ \frac{l(l+3)}{2(2l-1)} j_{l-2}[k(\eta_0-\eta)] + \frac{(l+1)(l-2)}{2(2l+3)} j_{l+1}[k(\eta_0-\eta)] \bigg\} \int_{k \eta_{\rm lss}}^{k \eta} dx \frac{-3 \, j_2(x)}{x} \frac{j_2(k \eta - x)}{(k \eta - x)^2}, \label{eq:FE}\end{align}

\noindent where the terms are defined as they were above for the $B_{lm}$ amplitudes. 

\subsection{Parity-Violating Correlations from Chiral GWs}

\begin{figure}[t!]
\centering
\includegraphics[width=0.8\textwidth]{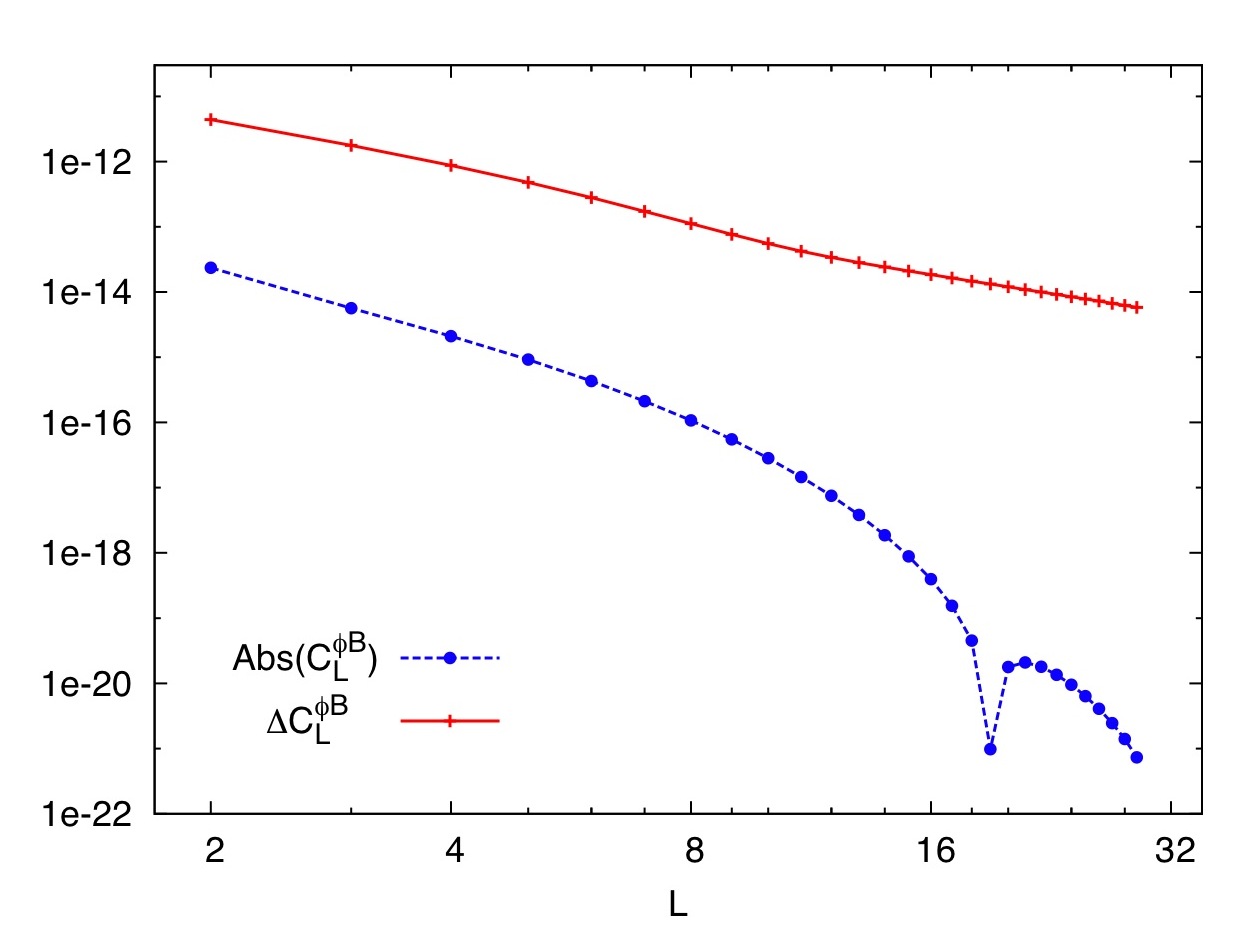}
\caption{Here we plot the cross-correlation $C_L^{\phi B}$ between the gradient $\phi$ modes of the weak lensing of cosmic shear with the curl-type $B$ modes of the CMB polarization in blue circles, and the noise on this measurement due to cosmic variance and Planck satellite instrumental noise in red $+$s. Since these quantities are of opposite parity, in the absence of parity-breaking physics we expect this cross-correlation to vanish. However, if we assume for example that the entire allowable GW background is right-circularly polarized, such a cross-correlation could occur. The cross-correlation is linearly proportional to the chirality parameter $A$, defined such that $A=1$ denotes a completely right-circularly polarized GW background, $A=-1$ denotes completely left-circularly polarized, and $A=0$ denotes an unpolarized background. Here we assume the maximum allowable tensor-to-scalar ratio $r=0.24$, the limit from WMAP-7 data combined with BAO and the $H_0$ measurement \cite{Komatsu:2010fb}. Cusps in the absolute value of the correlation function correspond to sign changes of the correlation function. }
\label{fig:phiB}
\end{figure}

\begin{figure}[t!]
\centering
\includegraphics[width=0.8\textwidth]{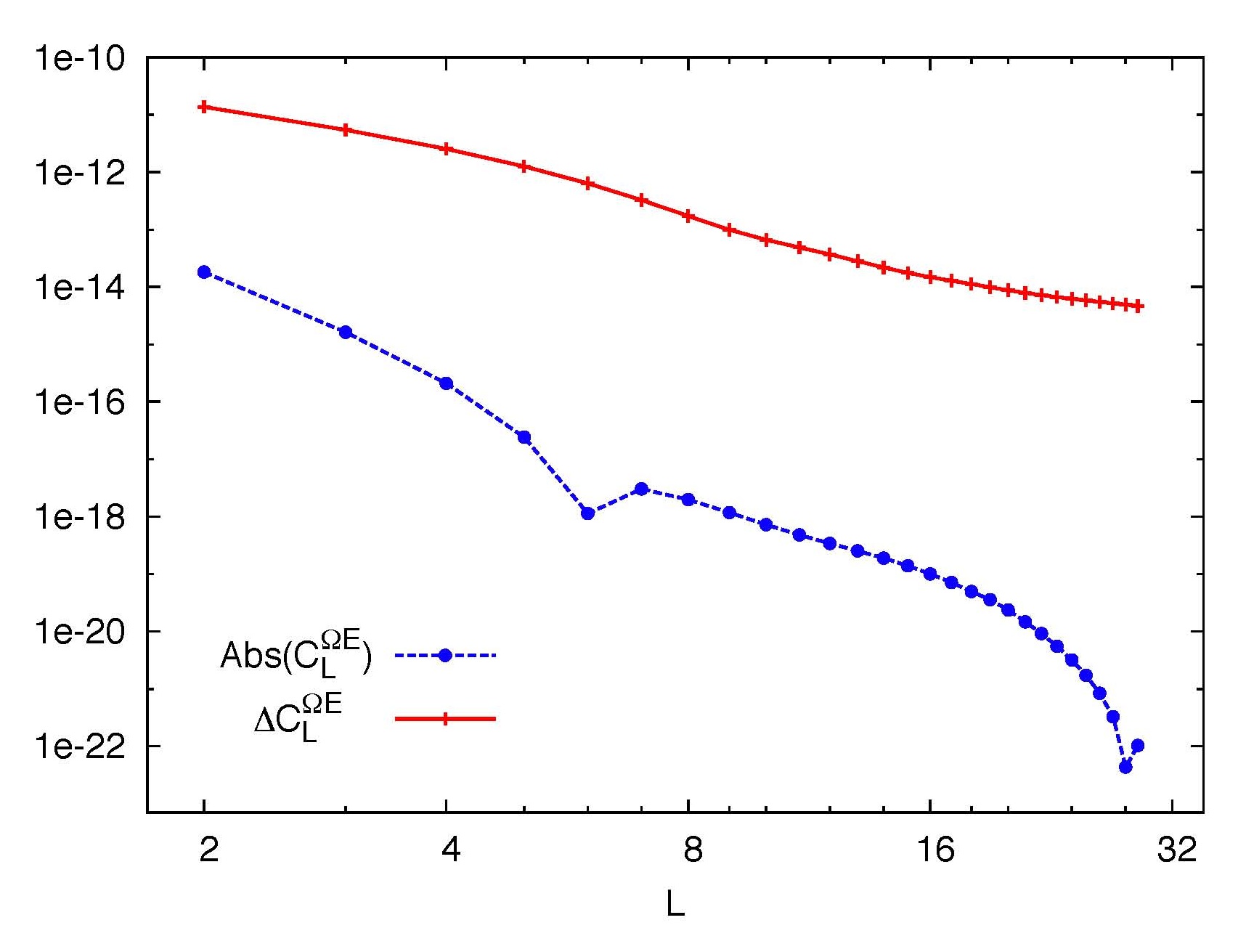}
\caption{Here we plot the cross-correlation $C_L^{\Omega E}$ between the curl-type $\Omega$ modes of cosmic shear with the gradient-type $E$-modes of the CMB polarization in blue circles, and the noise on this measurement due to cosmic variance and Planck satellite instrumental noise in red $+$s. As with the $\phi-B$ correlation, we assume a completely right-circularly polarized GW background, with the maximum currently permitted tensor-to-scalar ratio. }
\label{fig:OmegaE}
\end{figure}

We now want to calculate the expected cross-correlation between CMB-polarization multipole coefficients and weak-lensing-induced BiPoSHs of opposite parity. Note that these cross-correlations are directly related to the parity-odd three-point correlations discussed in Ref.~\cite{Kamionkowski:2010rb}. As we mentioned above, if there is no parity-violating physics, then in the cross-correlation of a parity-even and a parity-odd observable, $M=2$ terms and $M=-2$ terms will cancel each other, giving a net zero cross-correlation. However, if for example the GW background is chiral, then parity is broken and we can get a non-zero cross-correlation between opposite parity observables. As we saw in Sec.~\ref{sec:parity}, a right-handed GW contributes only to $M=2$ modes, while a left-handed one contributes only to $M=-2$. If we carry out a similar procedure for Eqs.~(\ref{eq:Bplkz}), and (\ref{eq:Eplkz}) as we did in Eq.~(\ref{eq:OmegaAadj}), weighting $M=2$ components by $(A+1)/2$ and $M=-2$ components by $(A-1)/2$, we can calculate parity-violating correlations between polarization and lensing components while accounting for the amplitude and handedness of a chiral GW background.

First considering the cross-correlation between $B$-modes of the CMB polarization and gradient-type modes of cosmic shear, we write 

\be C_L^{\phi B} = \frac{1}{2L+1} \sum_M \langle \phi_{LM} B^*_{LM} \rangle. \nonumber\ee

\noindent As before, by rotational invariance we know that both $+$ and $\times$ polarizations will contribute equally to $C_L^{\phi B}$, as will modes with any wavenumber $\vec{k}$ whose magnitude $k$ is the same. We can see that only $\phi_{LM}^{\rm ten}$ will contribute to this correlation, and not $\phi_{LM}^{\rm sca}$, as the scalar perturbation field is not correlated, on average, with the tensor perturbation field. Then using Eqs.~(\ref{eq:phimodeexp}), (\ref{eqn:phiresult}), (\ref{eqn:GWpowerspectrum}), (\ref{eq:Bmodeexp}) and (\ref{eq:Bplkz}), we can write this cross-correlation as

\be  C_L^{\phi B} = A \int \frac{d^3 k}{(2 \pi)^3} \: P_T(k) \: F^{\phi}_L(k) \: F^B_L(k). \label{eq:CphiB}\ee

Similarly, we can write the cross-correlation between $E$-type polarization modes and curl-type modes of cosmic shear, using Eqs.~(\ref{eq:Ommodeexp}), (\ref{eqn:Omegaresult}), (\ref{eqn:GWpowerspectrum}), (\ref{eq:Emodeexp}), and (\ref{eq:Eplkz}), as

\be C_L^{\Omega E} = A \int \frac{d^3 k}{(2 \pi)^3} \: P_T(k) \: F^{\Omega}_L(k) \: F^E_L(k), \label{eq:COmegaE}\ee

\noindent where the GW power spectrum is given by 

\be P_T(k) = \frac{\pi^2 \, r \: \Delta_R^2(k_0)}{2 \: k^3}. \nonumber\ee

We want to calculate the magnitude and shape of such
correlations, to determine whether such a signal is
observable. We use the WMAP 7-year cosmological parameters and
assume the maximum allowable level of GWs from early universe
physics, with a tensor-to-scalar ratio $r = 0.24$, the limit
from the WMAP-7 data combined with BAO and the $H_0$ measurement
\cite{Komatsu:2010fb}. We also assume that the GW background is
entirely right-circularly polarized. As a first estimate, we
calculate the level of such correlations while making several
assumptions. We use the approximate form of the GW transfer
function $T(k,\eta) \simeq 3 j_1(k\eta)/(k\eta)$, assume that
reionization happened instantaneously so that the electron
density $n_e$ is equal to a step function, and neglect
contributions to the polarization modes that came from last
scattering. The two last assumptions affect mostly the
higher-$L$ multipoles, which in this cross-correlation are
suppressed since we see that $\phi^{\rm ten}_{LM}$ and
$\Omega^{\rm ten}_{LM}$ fall off very fast with $L$.

With these assumptions, we have calculated the correlation functions $C_L^{\phi B}$ and $C_L^{\Omega E}$, and show them as the blue circles in Figs.~\ref{fig:phiB} and \ref{fig:OmegaE}. Note that the absolute value of the correlation functions are plotted, and that the cusps in the profiles result from sign changes. Note also that both correlation functions are linearly proportional to the chirality parameter $A$, so that they would flip in sign if the GW background were left instead of right-circularly polarized. We are only interested in low multipoles, since our assumptions break down for larger $L$, and such multipoles are strongly suppressed in correlation with the weak-lensing modes.

\subsection{Variance of $\phi$-B and $\Omega$-E Correlations}

It is useful to know the variance with which we could measure such parity-violating cross-correlations. From Ref.~\cite{Kamionkowski:1996ks} we see that the variance with which we could measure the cross-correlation $C_L^{XY}$ of two distinct Gaussian random variables $X$ and $Y$ is given by

\be \left( \Delta C_L^{XY} \right)^2 \equiv \left\langle \left( \widehat{C_L^{XY}} - C_L^{XY} \right)^2 \right\rangle, \nonumber\ee

\noindent where $\widehat{C_L^{XY}} = 1/(2L+1) \sum_M X_{LM} Y^*_{LM}$ is the estimator for the cross-correlation, and $C_L^{XY}$ is its theoretical value under the null hypothesis. Ref.~\cite{Kamionkowski:1996ks} then evaluates this variance, assuming distinct $X$ and $Y$, to be

\be \left( \Delta C_L^{XY} \right)^2 = \frac{1}{2L+1} \left[ \left( C_L^{XY} \right)^2 + C_L^{XX\,\rm map} \: C_L^{YY\,\rm map} \right], \label{eq:diagcov}\ee

\noindent where, as before, $C_L^{XX\,\rm map} = W_L^2 C_L + N_L^{XX}$, with $W_L$ the window function defined in Sec.~\ref{sec:oddbiposh}, and $N_L^{XX}$ the noise in the measurement of $C_L^{XX}$.

In our case, the null hypothesis is that there is a GW background with the maximal tensor-to-scalar ratio, but it contains equal numbers of right- and left-circularly polarized GWs, i.e., it is not chiral. In this case, the theoretical value of parity-violating cross-correlations is zero, so that the first term in Eq.~(\ref{eq:diagcov}) vanishes. Then, assuming that $\widehat{\phi_{LM}}$ and $\widehat{\Omega_{LM}}$ are Gaussian random variables, a reasonable assumption since many uncorrelated noise processes are likely to contribute to this measured value, we find for the variances,

\begin{align} \Big( \Delta C_L^{\phi B} \Big)^2 &= \frac{1}{2L+1} \: C_L^{\phi\phi\,\rm map} \: C_L^{BB\,\rm map} \\
\Big( \Delta C_L^{\Omega E} \Big)^2 &= \frac{1}{2L+1} \: C_L^{\Omega\Omega\,\rm map} \: C_L^{EE\,\rm map}. \end{align}

To calculate these errors, we know that the instrumental errors on the polarization power spectra are given by

\be N_L^{EE} = N_L^{BB} = \frac{8 \pi ({\rm NET})^2}{t_{\rm obs} \sqrt{f_{\rm sky}}}. \nonumber\ee

\noindent We use the Planck-satellite parameters, as in Sec.~\ref{sec:GWlensing}. We also use the CMB anisotropy calculator CAMB to calculate the temperature and polarization power spectra including effects at the surface of last scatter \cite{camb}. The resulting errors are shown as red $+$s in Figs.~\ref{fig:phiB} and \ref{fig:OmegaE}. This noise, which combines instrumental and cosmic-variance sources, is at least an order of magnitude above the corresponding maximum signal level at low multipoles, and drops less rapidly with $l$ so that the low multipoles yield the highest signal-to-noise.

\subsection{Signal-to-Noise Ratio of Chiral GW Background Detection} \label{sec:SNratio}

We finally wish to calculate the achievable signal-to-noise of a measurement of the magnitude of such cross-correlations given our calculations of their shapes and variances. Such a measurement would tell us about the presence or absence of a chiral GW background, or of parity violation in the processes that caused departures from Gaussianity/SI in general. We can phrase the aim of this calculation as finding the error with which we could measure the chirality parameter $A$, which sets the amplitude of the cross-correlations relative to their maximum values in the case of a completely circularly polarized GW background, as in Eqs.~(\ref{eq:CphiB}) and (\ref{eq:COmegaE}). Let us calculate this for the case of the $\phi$-$B$ cross-correlation; the $\Omega$-$E$ case will be similar.

We define a new quantity $C_{L\,\rm max}^{\phi B}$, defined such that

\be C_L^{\phi B} = A \: C_{L\,\rm max}^{\phi B}. \nonumber\ee

\noindent If we assume that the instrumental noise on $C_L^{\phi B}$ is Gaussian, so that $\widehat{C_L^{\phi B}} \equiv W_L^{-2} \, C_L^{\phi B}$ is a random variable drawn from a Gaussian probability distribution with variance $\Big( \Delta C_L^{\phi B} \Big)^2$ and mean $A \: C_{L\,\rm max}^{\phi B}$, we can find the maximum-likelihood estimator for $A$ to be

\be \widehat{A} = \frac{\sum_L \, \widehat{C_L^{\phi B}} \: C_{L\,\rm max}^{\phi B} \: \left( \Delta C_L^{\phi B} \right)^{-2}}{\sum_L \, \left(C_{L\,\rm max}^{\phi B}\right)^2 \, \left( \Delta C_L^{\phi B} \right)^{-2}} .\nonumber\ee

\noindent Then, assuming that the instrumental noise is uncorrelated between different multipoles, the variance of this estimator is given by

\be \left\langle \widehat{A}^2 \right\rangle = \left[ \sum_L \, \left(C_{L\,\rm max}^{\phi B}\right)^2 \, \left( \Delta C_L^{\phi B} \right)^{-2} \right]^{-1}. \nonumber\ee

The maximum signal-to-noise with which we can measure this amplitude is given by

\be \left( \frac{S}{N} \right)^{\phi B}_{\rm max} = \frac{\widehat{A}_{\rm max}}{\sqrt{\Big\langle \widehat{A}^2 \Big\rangle}} = \left[ \sum_L \, \left(C_{L\,\rm max}^{\phi B}\right)^2 \, \left( \Delta C_L^{\phi B} \right)^{-2} \right]^{1/2}. \ee

\noindent The same method can be used to calculate the obtainable signal-to-noise from the $\Omega$-$E$ cross correlation, giving

\be \left( \frac{S}{N} \right)^{\Omega E}_{\rm max} = \left[ \sum_L \, \left(C_{L\,\rm max}^{\,\Omega E} \right)^2 \, \left( \Delta C_L^{\,\Omega E} \right)^{-2} \right]^{1/2}. \ee

Using the values of the cross-correlations and their errors calculated above, we find that the obtainable signal-to-noise from measurement of these cross-correlations is 0.002 for $C_L^{\Omega E}$ and $0.01$ for $C_L^{\phi B}$. These numbers are too small for us to have any reasonable expectation of detection using the Planck satellite. Recalculating the above errors assuming an ideal CMB experiment, with no instrumental noise and infinite resolution, the values of the signal-to-noise only change by a factor of two, indicating that this method is not likely to be a promising way to detect a chiral GW background.

\section{Conclusions}

BiPoSHs are a formalism to describe correlations between two different spherical-harmonic coefficients of the CMB temperature field, which can occur if the CMB temperature field is not exactly Gaussian or statistically isotropic. This paper introduces odd-parity BiPoSHs, a set of BiPoSHs that has not yet been studied, and details how they can be estimated from knowledge of the CMB temperature fluctuations. 

We calculate the even- and odd-parity BiPoSHs that are sourced
by gradient- and curl-type deflections of the CMB, respectively,
and from this we obtain estimators for these deflections in
terms of the BiPoSH coefficients. We show that lensing by scalar
metric perturbations causes only gradient-type deflections, and
thus only sources even-parity BiPoSHs. However, lensing by GWs
produces both gradient- and curl-type deflections and thus
sources both even- and odd-parity BiPoSHs. We calculate the
expected power spectra of deflections due to scalar and tensor
perturbations and their errors, and conclude that a reasonable signal-to-noise measurement of
the amplitude of the GW background cannot be obtained from these
autocorrelations even with the ideal CMB experiment, and thus from autocorrelations of the BiPoSH coefficients.

Although lensing by GWs produces both even- and odd-parity BiPoSHs, their opposite parity implies that they could not be correlated. However, in the presence of parity-violating physics, such as a chiral GW background, this parity argument breaks down and we might expect a correlation. We consider such a cross-correlation, and encourage its measurement even though the likelihood of observing a cosmological signal is low. 

A GW background also produces signals in the $E$- and $B$-type
CMB polarization spectra, which are of even and odd parity,
respectively. We consider the possibility that a chiral GW
background would produce cross-correlations between
opposite-parity components of lensing and polarization, and
calculate the expected magnitude and errors of such
cross-correlations. Although we find that the likelihood of
observing a cosmological signal is low, we encourage the
measurement of these cross-correlations since such a detection
would provide evidence of important systematic errors or even
new parity-breaking physics.

Here we have discussed BiPoSHs constructed from temperature
multipole moments only, but the formalism can be generalized to
include the polarization as well.  It may also be that inclusion
of the polarization improves the sensitivity to these
parity-breaking, and other, signals.  We plan to pursue this
analysis in future work.

Finally, we note that weak-lensing
distortions of distant galaxies can also be decomposed into curl
and gradient components
\cite{Stebbins:1996wx,Kamionkowski:1997mp}.  Similar tests for
parity violation can thus also be carried out with weak lensing
of galaxies.

\begin{acknowledgments} 
LGB thanks Esfandiar Alizadeh, Christopher Hirata and {\bf Fabian Schmidt} for useful discussions, and acknowledges the support of the NSF Graduate Research Fellowship Program. MK thanks the support of the Miller Institute for Basic Research
in Science and the hospitality of the Department of Physics at
the University of California, Berkeley, where part of this work was
completed.  This work was supported at Caltech by DoE
DE-FG03-92-ER40701, NASA NNX10AD04G, and the Gordon and Betty Moore
Foundation.  TS acknowledges support from Swarnajayanti grant, DST,
India and the visit to Caltech during which the work was carried out. 
\end{acknowledgments} 

\end{document}